\documentclass[sigconf]{acmart}
\AtBeginDocument{%
  }

\setcopyright{acmlicensed}
\copyrightyear{2018}
\acmYear{2018}
\acmDOI{XXXXXXX.XXXXXXX}
\acmConference[Conference acronym 'XX]{Make sure to enter the correct
  conference title from your rights confirmation email}{June 03--05,
  2018}{Woodstock, NY}
\acmISBN{978-1-4503-XXXX-X/2018/06}

\usepackage{algorithm}
\usepackage{placeins}
\usepackage{algpseudocode}
\usepackage{booktabs}
\usepackage{graphicx}
\usepackage{lscape}
\usepackage{multirow}
\usepackage{subcaption}
\usepackage{enumitem}
\usepackage{amsmath}
\usepackage{subcaption}
\algnewcommand\REQUIRE{\Require}
\algnewcommand\ENSURE{\Ensure}
\algnewcommand\STATE{\State}
\algnewcommand\STATEX{\Statex}
\algnewcommand\FOR{\For}
\algnewcommand\IF{\If}
\algnewcommand\FORALL{\ForAll}
\algnewcommand\ENDIF{\EndIf}
\algnewcommand\ENDFOR{\EndFor}
\algnewcommand\RETURN{\Return}
\algnewcommand\COMMENT{\Comment}
\algnewcommand\FORP{\Forp}
\algnewcommand\ENDFORP{\Endforp}
\algnewcommand\TO{\To}

\usepackage{amsmath}  % 提供 \makecell 命令（单元格内换行必备，缺少会直接报错）
\usepackage{array}  

\begin{document}

\title{SMES: Towards Scalable Multi-Task Recommendation via Expert Sparsity}

\author{Yukun Zhang$^{*}$, Si Dong$^{*}$, Xu Wang$^{*}$\authornotemark[1], Bo Chen, Qinglin Jia, \\
Shengzhe Wang, Jinlong Jiao, Runhan Li, Jiaqing Liu, \\
Chaoyi Ma$^{\dagger}$, Ruiming Tang$^{\dagger}$\authornotemark[2], Guorui Zhou, Han Li, Kun Gai}
% 修正原authors命令里的拼写错误（Jinglong→Jinlong，Ga→Gai）
\renewcommand{\authors}{Yukun Zhang, Si Dong, Xu Wang, Bo Chen, Qinglin Jia, Shengzhe Wang, Runhan Li, Jinlong Jiao, Jiaqing Liu, Chaoyi Ma, Ruiming Tang, Guorui Zhou, Han Li, Kun Gai}

% 核心修改：添加country字段（快手总部位于中国北京）
\affiliation{%
  \institution{Kuaishou Technology Co., Ltd.}  % 机构名称
  \city{Beijing}                               % 城市（必填，acmart推荐标注）
  \country{China}                              % 国家（解决编译错误的核心）
}

% 修正邮箱列表里的错误（原代码混入renze03/dukang05，且格式优化）
\email{{zhangyukun09, dongsi, wangxu28, renze03, dukang05, wangshengzhe, jiaojinlong, lirunhan}@kuaishou.com}
\email{{liujiaqing, machaoyi03, tangruiming, zhouguorui, lihan08}@kuaishou.com, gai.kun@qq.com}
\renewcommand{\shortauthors}{Zhang and Dong, et al.}
\begin{abstract}

Industrial recommender systems typically rely on multi-task learning to estimate diverse user feedback signals and aggregate them for ranking. 
Recent advances in model scaling have shown promising gains in recommendation.
However, naively increasing model capacity imposes prohibitive online inference costs and often yields diminishing returns for sparse tasks with skewed label distributions. This mismatch between uniform parameter scaling and heterogeneous task capacity demands poses a fundamental challenge for scalable multi-task recommendation.
In this work, we investigate parameter sparsification as a principled scaling paradigm and identify two critical obstacles when applying sparse Mixture-of-Experts (MoE) to multi-task recommendation: \textbf{exploded expert} activation that undermines instance-level sparsity and \textbf{expert load skew} caused by independent task-wise routing. To address these challenges, we propose SMES, a scalable sparse MoE framework with \textbf{progressive expert routing}. SMES decomposes expert activation into a task-shared expert subset jointly selected across tasks and task-adaptive private experts, explicitly bounding per-instance expert execution while preserving task-specific capacity. In addition, SMES introduces a global \textbf{ multi-gate load-balancing regularizer} that stabilizes training by regulating aggregated expert utilization across all tasks.
SMES has been deployed in Kuaishou’s large-scale short-video services, supporting over 400 million daily active users. Extensive online experiments demonstrate stable improvements, with GAUC gain of \textbf{0.29\%} and a \textbf{0.31\%} uplift in user watch time.

 % achieving superior efficiency–capacity trade-offs under strict latency constraints. 

\end{abstract}




\maketitle

\section{INTRODUCTION}
Industrial recommender systems infer user preferences by modeling users’ historical interaction behaviors to deliver personalized recommendations, which have been widely deployed in platforms such as e-commerce, short-video services, and online advertising~\cite{autodis,sim,rankmixer}. Owing to the lack of explicit user preference signals, practical recommender systems typically rely on multi-task learning frameworks~\cite{mmoe,ple,taml,nooneleft}
to predict diverse dimensions of users’ implicit feedback. For example, in short-video recommendation scenarios like Kuaishou and TikTok~\cite{home,longer}, recommender systems often simultaneously predict dozens or even hundreds of interaction probabilities, such as \textit{Click, Like, Share, Comment, Long-view}, and \textit{Completion}. To estimate users’ overall satisfaction with candidate items, these predictive probabilities are combined subsequently through aggregation functions or models~\cite{xmtf,pantheon} to produce a final relevance or utility score for ranking.

\begin{figure}[t] 
    \centering  
    % 子图1：Valid_play（宽度从0.32\hsize→0.31\hsize，放大图片占比）
    \begin{subfigure}{0.32\hsize}  % 减少子图容器宽度预留间距
        \centering
        \includegraphics[width=1.0\hsize]{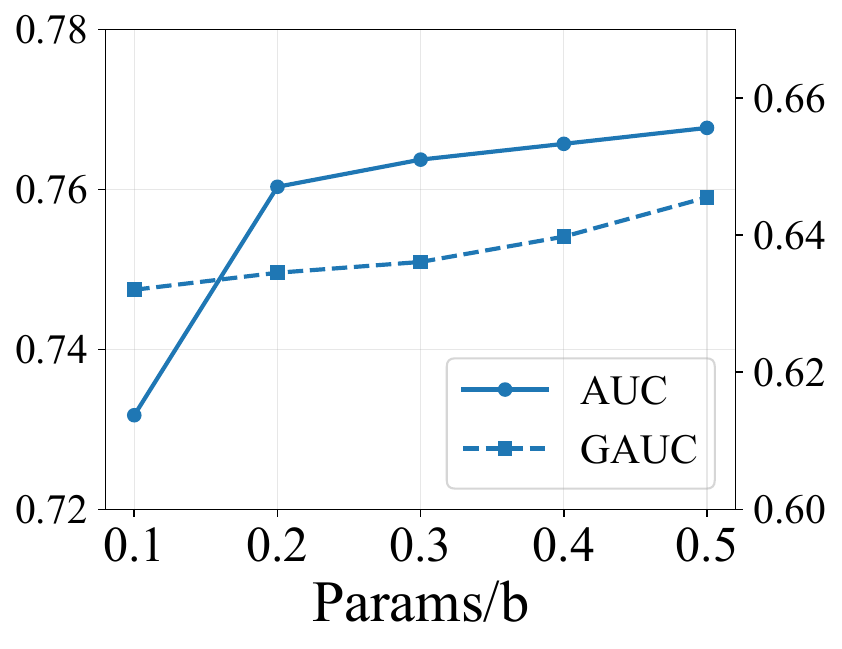}  % 图片占满子图容器（0.99→1.0）
        \caption{Effective-view}
        \label{subfig:Effective-view}
    \end{subfigure}
    \hfill  % 子图间水平间距（从自动填充改为固定1pt，大幅缩小）
    % 子图2：Like
    \begin{subfigure}{0.32\hsize}
        \centering
        \includegraphics[width=1.0\hsize]{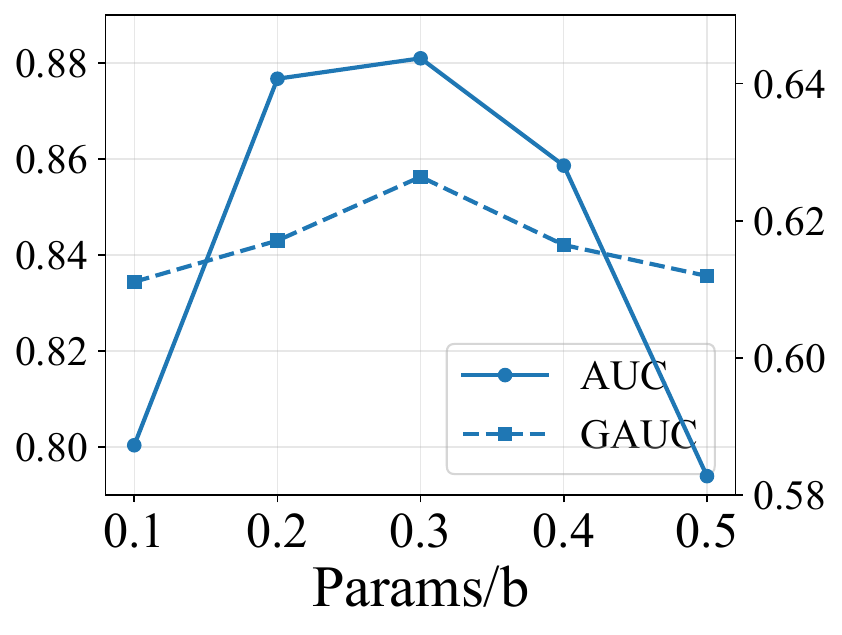}
        \caption{Like}
        \label{subfig:like}
    \end{subfigure}
    \hfill  % 统一间距
    % 子图3：Follow
    \begin{subfigure}{0.32\hsize}
        \centering
        \includegraphics[width=1.0\hsize]{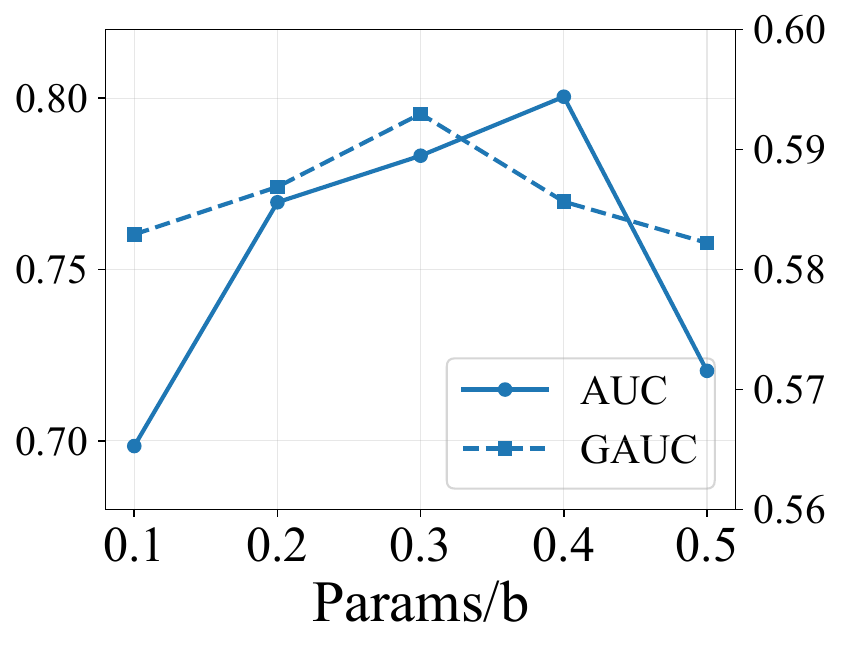}
        \caption{Follow}
        \label{subfig:follow}
    \end{subfigure}
    
    % 整体标题（保持原有内容，优化换行避免溢出）
    \caption{\textbf{Performance trends across tasks in KuaiRand dataset.} 
        The curves show AUC (solid line) and GAUC (dashed line) trends with increasing model parameter scale (billion). For tasks (e.g., \textit{watching-time task,} \textit{Effective-view}), performance consistently improves with increasing parameter scale. In contrast, for tasks (e.g., \textit{interaction tasks} such as \textit{like} and \textit{follow}), blindly increasing model capacity can lead to performance degradation.}
    \label{Trend_of_different_task}
    \vspace{-6pt}  % 减少图片下方空白
\end{figure}

Benefiting from advances in computational resources and the substantial gains brought by model scaling in the large language model (LLM) community~\cite{llama,qwen3}, many industrial recommender platforms have recently explored scaling approaches tailored to recommendation scenarios and achieved promising results (e.g., HSTU~\cite{hstu} and RankMixer~\cite{rankmixer}). However, industrial recommender systems operate under strict constraints of high concurrency and low latency, with stringent requirements on online inference~\cite{autogen}. Pursuing performance gains through parameter-scaling alone imposes significant burdens on online serving, ultimately harming system-level return on investment (ROI). Moreover, tasks across different data scale and sparsity regimes exhibit fundamentally distinct capacity demands~\cite{ple,home}. 
Figure~\ref{Trend_of_different_task} illustrates the performance trends of different tasks on the public KuaiRand dataset\cite{gao2022kuairand}, as the model parameter scale increases. The results show that for some tasks (e.g., \textit{like} and \textit{follow}), scaling up model parameters does not lead to consistent performance improvements.
The existing scaling methods~\cite{hstu,rankmixer} often overlook the characteristics, and this misalignment between uniform capacity scaling and task-specific supervision characteristics motivates adaptive scaling strategies for multi-task recommendation scenarios.

Therefore, we explore a new parameter scaling paradigm for multi-task recommendation, which simultaneously balances the performance gains from model scaling with stringent latency constraints and accommodates the heterogeneous capacity requirements of different tasks.
Our core solution is parameter sparsification. On the one hand, sparse activation during online inference effectively alleviates latency overhead and reduces the computational cost induced by model scaling. On the other hand, sparsification enables different tasks to adaptively leverage an appropriate amount of model capacity, thereby avoiding unstable generalization caused by over-parameterization for sparse tasks.

Despite its appealing efficiency–capacity trade-off, directly applying sparse Mixture-of-Experts (MoE) to multi-task recommendation is non-trivial.
Existing MoE systems typically balance experts per router or per task, which is sufficient in single-task settings but inadequate under multi-task sparse routing, where expert usage is aggregated across tasks at the instance level.
MTL introduces multiple task-specific routers whose sparse decisions jointly determine the expert computation at inference time.
In practice, naive task-wise sparse routing suffers from two fundamental issues.
First, independent top-$k$ routing across tasks leads to \textbf{exploded expert activation}. Different tasks may activate largely disjoint expert subsets for the same instance, causing the union of activated experts to grow rapidly with the number of tasks and undermining instance-level sparsity.
Conversely, excessive overlap may collapse multiple tasks onto a small subset of experts, limiting effective capacity and task specialization.
Second, sparse routing exacerbates \textbf{severe expert load skew} under MTL. Since expert updates are triggered only when selected, traffic and gradients from multiple tasks can accumulate on a few popular experts, while many others remain rarely activated and under-trained. Such imbalance becomes increasingly severe as the number of tasks grows, leading to unstable training dynamics and inefficient utilization of large expert pools.

To address these challenges, we propose SMES (Scalable Multi-task recommendation via Expert Sparsity), a sparse MoE framework that translates parameter scaling into consistent performance gains in multi-task recommendation.
SMES introduces a \textbf{progressive expert routing mechanism} that decomposes expert activation into an instance-level shared expert subset and a task-adaptive expert subset.
By selecting a small set of shared experts jointly preferred by all tasks and allowing each task to activate a limited number of additional private experts, SMES explicitly controls exploded expert activation while preserving task-specific capacity. 
This design bounds the number of distinct experts executed per instance, enabling predictable and scalable inference under strict latency budgets. 
Furthermore, SMES employs a \textbf{multi-task load-balancing regularizer} that operates on aggregated expert usage across all tasks, rather than balancing each task router independently. This global regularization mitigates expert hotspots induced by multi-task sparse routing and stabilizes training as the expert pool scales. We also incorporate deployment-specific optimizations to enable efficient execution and maintain low-latency serving at scale.

Overall, the contributions are mainly summarized as follows:
\begin{itemize}
\item We identify and analyze the unique challenges of parameter scaling with sparse MoE in multi-task recommendation, highlighting the tension between scaling model capacity, task specialization, and aggregated expert utilization.
\item We propose SMES, a sparse multi-gate MoE framework that enables efficient parameter scaling in multi-task recommendation via coordinated routing and cross-task load balancing.
\item Extensive experiments on public benchmarks and large-scale industrial datasets demonstrate that SMES consistently outperforms classic dense and naive sparse baselines, achieving superior efficiency–capacity trade-offs under strict online serving constraints.
\item We further develop an efficient online deployment and execution strategy for SMES, including optimized matrix computation and memory-efficient execution, supporting large-scale low-latency production serving on Kuaishou and yielding a 0.31\% lift in user watch time. 
\end{itemize}

\section{RELATED WORK}
This section briefly reviews the optimization research of multi-task learning (MTL) and model scaling in recommender systems. 

\subsection{Multi-Task Recommendation}
% This section briefly reviews the optimization research of multi-task learning (MTL) in recommender systems. MTL plays a key role in recommender systems, neural language processing and other fields []. \textbf{}
For industrial recommendation, it is essential to predict multiple tasks simultaneously, such as Click, Like, Share, and other user interactions. Therefore, multi-task learning has been a key area of focus in industrial research for a long time.

Early methods relied on hard expert architectures such as shared bottom~\cite{sharebottom} and mixture-of-experts (MoE)~\cite{moe}, which were difficult to adapt to complex task relationships. Multi-gate Mixture-of-Experts (MMoE)~\cite{mmoe} applied the MoE structure to multi-task learning and explicitly modeled task relationships through shared experts and task-specific gating, which performed better in low-relevance scenarios.
To alleviate the seesaw phenomenon, progressive layer extraction (PLE)~\cite{ple} separated shared and task-specific components and improved joint representation efficiency through progressive routing. Adaptive Task-to-Task Fusion Network (AdaTT)~\cite{adatt} further introduced residual and gating mechanisms to adaptively learn shared and task-specific knowledge via fusion units. Moreover, the Multi-Level Sparse Sharing Model (MSSM)~\cite{mssm} designed field-level and unit-level sparse modules to address negative transfer caused by indistinguishable feature sharing.
To tackle the parameter and resource bottlenecks of MoE-based methods, Mixture-of-Masked-Experts (MoME)~\cite{mome} extracted experts from a base network using binary masks and reduced storage requirements through coarse and fine-grained pruning. Hierarchy of Multi-Gate Experts (HoME)~\cite{home} adopted expert normalization, hierarchical masking, and feature gating mechanisms to mitigate expert collapse, degradation, and underfitting.
% respectively.

\subsection{Model Scaling}
Due to the significant performance improvements brought by scaling up in the LLM field, recommendation models have also explored scaling strategies in recent years.
Wukong~\cite{wukong} proposed a stacked factorization machine architecture and a collaborative scaling strategy, capturing arbitrary sequence feature interactions by increasing the number of layers and model width.
HSTU~\cite{hstu} organized samples in a user-centric manner and designed a Hierarchical Sequential Transduction Unit to model sequential information, exploring the scaling of recommendation models through generative training approaches.
HHFT~\cite{hhft} improved the model scaling scheme for large-scale heterogeneous feature scenarios through stacked heterogeneous Transformer and Hiformer~\cite{hiformer} layers.
Moreover, OneTrans~\cite{onetrans} encoded sequential and non-sequential features into a unified token space and leveraged a shared Transformer block to jointly perform sequence modeling and feature interaction.
Beyond the Transformer-based scaling strategies discussed above, RankMixer~\cite{rankmixer} explored a more efficient modeling paradigm by replacing self-attention with a multi-head token-mixing module combined with per-token feed-forward networks (FFNs) to enable efficient feature interaction, thereby scaling the model to one billion parameters. In addition, it adopted a sparse MoE architecture to improve overall ROI and alleviate expert training imbalance.

%To address the limitations of independent optimization in existing works, OneTrans~\cite{onetrans} achieves simultaneous optimization of feature interaction and sequence modeling by fusing sequential and non-sequence tokens through a unified tokenizer. To better scaling up model, it stacked onetrans modules which share parameters between similar sequence tokens and assign token-specific parameters to non-sequence tokens.

Despite the impressive results achieved by current model scaling approaches, they still face several challenges in industrial deployment.
Industrial recommendation is constrained by high concurrency and low latency, and simply scaling parameters can easily increase service burden and reduce ROI. Furthermore, existing methods do not adapt to the heterogeneous capacity requirements of tasks with different sparsity, leading to suboptimal performance. Therefore, it is crucial to explore a new paradigm for scaling multi-task recommendation parameters that balances latency and performance while adapting to heterogeneous task capacities.

\section{METHODOLOGY}

\subsection{Preliminaries}
\label{sec:mtl}
\begin{figure*}[!htbp]
	\centering
	\includegraphics[width=0.98\textwidth]{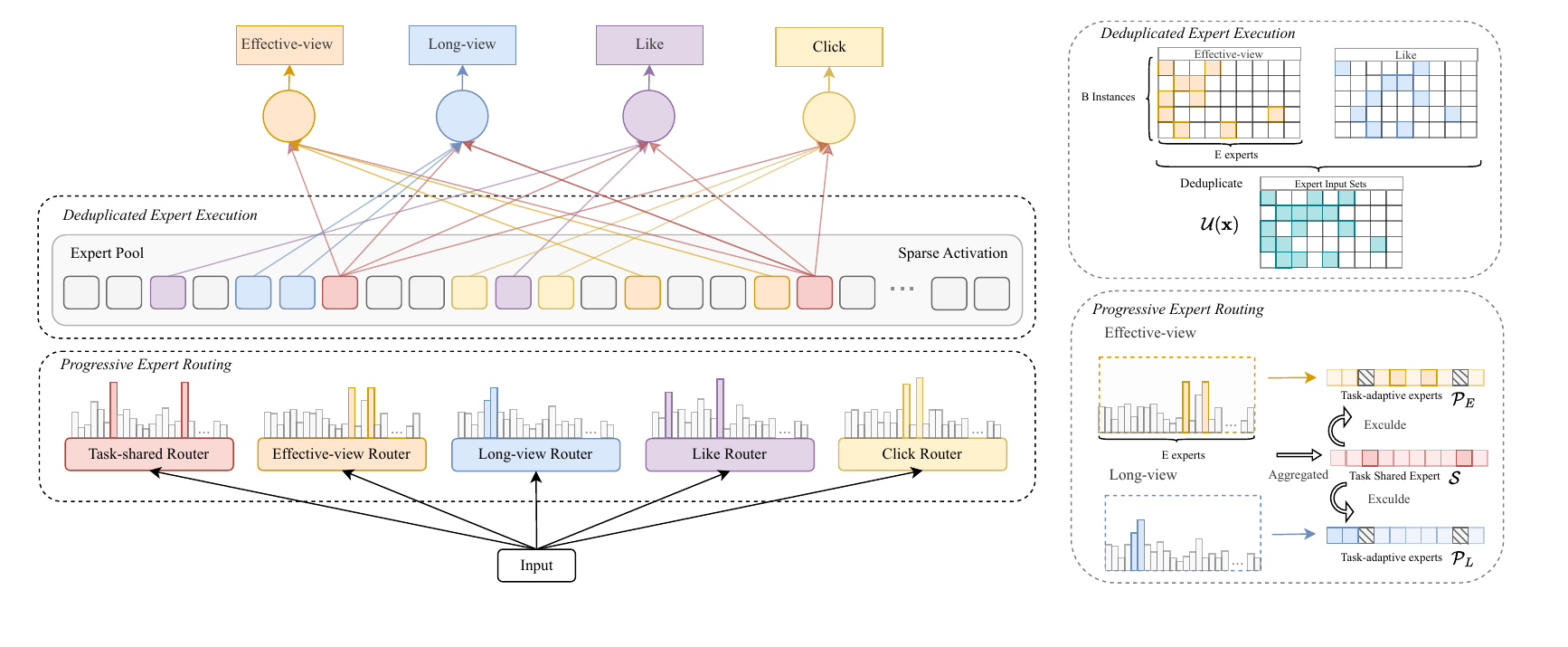}
	\caption{\textbf{The architecture of our SMES for multi-task recommendation.} 
	It contains two key components: 
	(1) Progressive Expert Routing uses a task-shared router and task-adaptive sub-routers to select experts 
	(only Effective-view and Long-view tasks are visualized for clarity). 
	(2) Deduplicated Expert Execution removes redundant expert computations across tasks.}
	\label{fig:method}
	\vspace{-6pt}
\end{figure*}

We consider a multi-task recommendation model trained on $T$ prediction tasks $\mathcal{T}=\{1,\ldots,T\}$ (e.g., click, effective-view, like). Each training instance is represented by an input feature vector $\mathbf{x}\in\mathbb{R}^{d}$ that concatenates user, item, and contextual features~\cite{edcn}, paired with the associated labels $\mathbf{y}=\{y_t\}_{t=1}^{T}$.

Specifically, a shared encoder $F(\cdot)$ maps the input $\mathbf{x}$ into a dense representation, capturing intrinsic user-item-context interactions. This encoder is typically instantiated as embedding layers followed by a deep representation learning backbone, such as DeepFM~\cite{deepfm}, DCN~\cite{dcn}, DIN~\cite{din}, or SIM~\cite{sim}:
\begin{equation}
\mathbf{h}=F(\mathbf{x})\in\mathbb{R}^{d_{\mathrm{in}}}.
\end{equation}
Subsequently, a task-specific prediction head $\phi_t(\cdot)$ (e.g., Multi-Layer Perceptron) projects the shared representation $\mathbf{h}$ to the prediction for task $t$:
\begin{equation}
\hat{y}_t=\phi_t(\mathbf{h}).
\end{equation}

The overall training objective is a weighted sum of task losses:
\begin{equation}
\mathcal{L}_{\mathrm{total}}=\sum_{t=1}^T \lambda_t\,\mathcal{L}_t(y_t,\hat{y}_t),
\end{equation}
where $\lambda_t \ge 0$ serves as a hyper-parameter balancing the contribution of task $t$, and $\mathcal{L}_t$ denotes the task-specific loss function (e.g., binary cross-entropy).

\subsubsection{Dense Multi-Task MoE and Scaling Bottlenecks}
\label{sec:dense_mmoe}

% TODO 全文的gate和router
A common baseline directly stacks task-specific heads on the shared representation, formulated as $\hat{y}_t=\phi_t(\mathbf{h})$. However, when tasks are heterogeneous (e.g., possessing distinct decision boundaries or generating partially conflicting gradients), forcing them to share the exact same feature subspace can restrict task specialization and induce negative transfer~\cite{zhang2021survey}.
To facilitate \emph{task-adaptive} feature sharing, 
 % Multi-gate Mixture-of-Experts 
MMoE~\cite{mmoe} introduces a mixture-of-experts layer on top of $\mathbf{h}$. This architecture maintains a set of $E$ experts $\{f_e\}_{e=1}^{E}$ and assigns a dedicated gate (router) $g_t$ to each task $t$. Specifically, each expert $f_e:\mathbb{R}^{d_{\mathrm{in}}}\rightarrow\mathbb{R}^{d_{\mathrm{out}}}$ transforms the shared representation into an expert output $\mathbf{o}_e$, while each gate $g_t:\mathbb{R}^{d_{\mathrm{in}}}\rightarrow\mathbb{R}^{E}$ generates routing logits $\mathbf{z}_t$ over the experts. Concretely,
\begin{equation}
\begin{aligned}
\mathbf{o}_e &= f_e(\mathbf{h}) \in \mathbb{R}^{d_{\mathrm{out}}}, \qquad e=1,\ldots,E,\\
\mathbf{z}_t &= g_t(\mathbf{h}) \in \mathbb{R}^{E},\\
\mathbf{p}_t &= \mathrm{softmax}(\mathbf{z}_t) \in \mathbb{R}^{E},\\
\mathbf{h}_t &= \sum_{e=1}^{E} p_{t,e}\,\mathbf{o}_e \in \mathbb{R}^{d_{\mathrm{out}}},
\end{aligned}
\label{eq:mmoe}
\end{equation}
where $\mathbf{p}_t$ lies on the probability simplex ($p_{t,e}\!\ge\!0$ and $\sum_{e=1}^{E}p_{t,e}\!=\!1$), and $p_{t,e}$ denotes the $e$-th entry of $\mathbf{p}_t$. The task-specific representation $\mathbf{h}_t$ is then fed into the corresponding head:
$\hat{y}_t=\phi_t(\mathbf{h}_t)$.

In principle, MoE capacity can be increased by scaling the number of experts $E$, thereby increasing parameters and allowing different experts to capture diverse patterns in recommendation data. However, most recommender deployments adopt \textbf{expert-dense} MoE variants, where all $E$ experts are computed and then aggregated for each task. Consequently, both expert-side computation and intermediate activations scale linearly with $E$:
\begin{equation}
\text{FLOPs/instance}\ \propto\ E,\qquad
\text{ActMem/instance}\ \propto\ E,
\end{equation}
which conflicts with strict online latency and memory budgets.

Moreover, tasks operating at varying data scales exhibit fundamentally distinct capacity demands~\cite{ple,home}. In industrial multi-task recommendation scenarios, tasks often encounter highly heterogeneous data regimes. While some tasks benefit from abundant supervision signals, others suffer from severe label sparsity and skewed distributions. This misalignment between static, dense model capacity and diverse task complexities inevitably leads to significant resource wastage and suboptimal serving efficiency.

\subsubsection{Sparse MoE and MTL-Specific Challenges}
\label{sec:sparse_moe_mtl}

To reconcile the conflict between capacity scaling and strict online latency budgets, and to adaptively match model capacity with diverse task complexities, we adopt \emph{sparse routing}. This paradigm decouples the parameter size from the active computation per instance.
In a standard single-task scenario, sparse MoE dynamically selects a subset of experts via a routing mechanism~\cite{sparsemoe,switch}.
Given router logits $\mathbf{z}\in\mathbb{R}^{E}$ over $E$ experts, the set of activated indices is determined by:
\begin{equation}
\mathcal{K} = \mathrm{TopK}\big(\mathbf{z}, K\big),
\quad K\ll E,
\end{equation}
where $\mathcal{K}\subset\{1,\dots,E\}$ denotes the index set of the $K$ selected experts with the highest routing scores, and $K$ denotes the routing budget. The \emph{routing weights} are then computed by normalizing the probability mass strictly within the selected experts:
\begin{equation}
p_e=
\frac{\exp(z_e)}{\sum_{j\in \mathcal{K}}\exp(z_j)}
\cdot \mathbb{I}\!\left[e\in\mathcal{K}\right],
\label{eq:sparse_p}
\end{equation}
where $\mathbb{I}[\cdot]$ is the indicator function. The output is computed as $\tilde{\mathbf{h}}=\sum_{e\in\mathcal{K}} p_e\,\mathbf{o}_e$. Compared to dense MoE, sparse routing reduces the computational complexity from $\mathcal{O}(E)$ to $\mathcal{O}(K)$ per instance, theoretically enabling massive scaling of $E$.

\paragraph{Naive multi-task sparse routing}
A straightforward extension to multi-task learning is to equip each task $t$ with an independent router $g_t$ and perform task-wise independent sparse routing. For an instance $i_g$, task $t$ selects its top-$K$ experts $\mathcal{K}_t=\mathrm{TopK}(\mathbf{z}_t,K)$. However, in industrial recommendation where the number of tasks $T$ is large, this naive design faces two critical challenges:

\begin{itemize}[leftmargin=*, topsep=2pt, itemsep=4pt]
    \item \textbf{Exploded Expert Activation.}
    While each task respects the budget $K$, the system-level computational cost for processing instance $i_g$ depends on the \emph{union} of activated experts across all tasks, denoted as $\mathcal{U}=\bigcup_{t=1}^{T}\mathcal{K}_t$. The effective number of active experts is bounded by:
    \begin{equation}
    K \;\le\; |\mathcal{U}| \;\le\; \min\{E,TK\}.
    \end{equation}
    When tasks select disjoint sets of experts due to task diversity, $|\mathcal{U}|$ grows with $T$. Given that $TK$ can easily exceed $E$ in practice, this ``union effect'' causes the model to degenerate back into a dense-like regime(activating nearly all experts).
    % nullifying the efficiency gains of sparse routing. 
    By contrast, activating only $K$ experts ($K \ll E$) can collapse tasks onto a small expert subset, limiting effective capacity.

\item \textbf{Severe Expert Load Skew.}
    For a mini-batch of size $B$, the cumulative routing load for expert $e$ is calculated by aggregating selections across all tasks and instances:
    \begin{equation}
    c_e=\sum_{b=1}^{B}\sum_{t=1}^T \mathbb{I}\!\left[e\in\mathcal{K}_t^{(b)}\right].
    \end{equation}
    As the load is aggregated across $T$ independent routers, the resulting distribution $\{c_e\}_{e=1}^{E}$ tends to be severely skewed. This accumulation effect causes a small subset of experts to dominate traffic and gradients, while the majority remain underutilized and undertrained. 
    % Such imbalance wastes capacity as $E$ grows and significantly destabilizes training.
\end{itemize}

\subsection{SMES: Scalable Multi-Task Recommendation via Expert Sparsity}
\label{sec:SMES}

To overcome the computational bottlenecks of dense MMoE (Sec.~\ref{sec:dense_mmoe}) while resolving the \emph{exploded activation} and \emph{load skew} pathologies inherent to naive multi-task sparse routing (Sec.~\ref{sec:sparse_moe_mtl}), we propose \textbf{Scalable Multi-Task Recommendation via Expert Sparsity (SMES)}, as illustrated in Figure~\ref{fig:method}.

SMES retains the standard multi-task architecture (shared encoder $F$, expert pool $\{f_e\}_{e=1}^{E}$, and task heads $\{\phi_t\}_{t=1}^{T}$), while redesigning \emph{expert routing and execution}. Specifically, SMES integrates two key components: \textbf{Progressive Expert Routing (PER)}, which strictly bounds the number of \textit{distinct} experts executed per instance to prevent activation explosion; and a \textbf{Multi-Task Load-Balancing (MTLB) Regularizer}, which mitigates cross-task hotspots to stabilize aggregated expert utilization.

\subsubsection{Progressive Expert Routing}
\label{sec:coord_routing}

\paragraph{Task Routers.}
Following Eq.~\eqref{eq:mmoe}, each task router $g_t$ maps the shared representation $\mathbf{h}$ to logits $\mathbf{z}_t \in \mathbb{R}^{E}$. These logits are then normalized to obtain scores $\mathbf{p}_t = \mathrm{softmax}(\mathbf{z}_t)$. Here, $z_{t,e}$ and $p_{t,e}$ denote the $e$-th entry of $\mathbf{z}_t$ and $\mathbf{p}_t$, respectively.

We generalize the concept of shared and specific experts to a dynamic, sparse routing context. For each instance $i_g$ per task, SMES activates only $K\ll E$ experts, decomposed as $K = K_s + K_a$, where $K_s$ experts form a \textbf{task-shared} set across all tasks, and $K_a$ experts are \textbf{task-adaptive} and selected separately for each task.

\paragraph{Stage-I: Task-shared experts.}
We compute a \textbf{global routing score} for each expert by weighted pooling \emph{normalized} task-specific routing probabilities:
\begin{equation}
s_e=\sum_{t=1}^{T} w_tp_{t,e},
\label{eq:agg_score}
\end{equation}
where $w_t\ge 0$ is an optional task weight that reflects task importance (default $w_t=1$).
The shared expert set is then selected as
\begin{equation}
\mathcal{S}=\mathrm{TopK}\left(\{s_e\}_{e=1}^{E},\,K_s\right),
\quad |\mathcal{S}|=K_s.
\label{eq:shared_set}
\end{equation}

\paragraph{Stage-II: Task-adaptive experts.}
For each task $t$, we select $K_a$ additional experts from the remaining candidates (complement of $\mathcal{S}$) based on the corresponding logits $z_{t,e}$:
\begin{equation}
    \mathcal{A}_t = \mathrm{TopK}\!\left(\{z_{t,e}\}_{e\notin \mathcal{S}},\,K_a\right),
    \quad |\mathcal{A}_t|=K_a.
    \label{eq:private_set}
\end{equation}
The final activated set for task $t$ is:
\begin{equation}
\mathcal{K}_t = \mathcal{S} \cup \mathcal{A}_t,
\quad |\mathcal{K}_t| = K.
\label{eq:active_set}
\end{equation}
By construction, all tasks share $\mathcal{S}$, guaranteeing a minimum overlap of $K_s$ experts per instance, while
$\mathcal{A}_t$ preserves task-adaptive capacity.

\paragraph{Deduplicated Expert Execution.}
\label{sec:union}
Given the activated sets $\{\mathcal{K}_t\}_{t=1}^{T}$, a naive sparse implementation executes experts independently for each task, redundantly recomputing $\mathbf{o}_e$ whenever multiple tasks activate the same expert. SMES removes this redundancy by executing each selected expert \emph{at most once per instance} on the cross-task set:
\begin{equation}
\mathcal{U} = \bigcup_{t=1}^{T} \mathcal{K}_t,
\quad |\mathcal{U}| \le K_s + T K_a.
\end{equation}
We then compute expert outputs $\mathbf{o}_e$ only for the experts $e\in\mathcal{U}$.
% \begin{equation}
% \mathbf{o}_e = f_e(\mathbf{h}), \qquad \forall e\in\mathcal{U}.
% \end{equation}
The resulting outputs $\{\mathbf{o}_e\}_{e\in\mathcal{U}}$ are shared across all tasks in the subsequent aggregation, reducing expert executions per instance from up to $TK$ to $|\mathcal{U}|$. % (often much smaller when tasks overlap through $\mathcal{S}$).

\paragraph{Sparse Aggregation.}
Given $\mathcal{K}_t$ and the computed expert outputs, we compute the final routing weights $p_{t,e}$ by renormalizing \emph{only} over activated experts:
\begin{equation}
p_{t,e}=
\frac{\exp\!\left(z_{t,e}\right)}{\sum_{j\in \mathcal{K}_t} \exp\!\left(z_{t,j}\right)}
\cdot \mathbb{I}\!\left[e\in\mathcal{K}_t\right].
\label{eq:sparse_weight}
\end{equation}
The task-specific representation is $\mathbf{h}_t=\sum_{e\in\mathcal{K}_t} p_{t,e}\,\mathbf{o}_e$,
followed by the task head $\hat{y}_t=\phi_t(\mathbf{h}_t)$.

\subsubsection{Multi-Task Load-Balancing Regularization}
\label{sec:balance}

To prevent \emph{cross-task} expert hotspots under multi-task top-$K$ routing, we regularize expert utilization by
balancing the \emph{aggregate} traffic induced by \emph{all} task routers.
For a mini-batch of size $B$, we define the \emph{selection frequency} $\bar f_e$ and the \emph{probability mass} $\bar p_e$ for expert $e$ as
\begin{equation}
\begin{aligned}
\bar f_e &= \frac{1}{BT}\sum_{b=1}^{B}\sum_{t=1}^{T}
\mathbb{I}\!\left[e\in \mathcal{K}_t^{(b)}\right],\\
\bar p_e &= \frac{1}{BT}\sum_{b=1}^{B}\sum_{t=1}^{T}
p_{t,e}^{(b)}.
\end{aligned}
\end{equation}
where $\mathcal{K}_t^{(b)}$ denotes the experts activated by task $t$ for the $b$-th instance in the mini-batch. We then minimize
\begin{equation}
\mathcal{L}_{\mathrm{lb}}=\frac{E}{K}\sum_{e=1}^{E} \bar f_e\, \bar p_e,
\end{equation}
which discourages concentrating \emph{both} high realized traffic ($\bar f_e$) and high routing preference ($\bar p_e$) on the same experts \emph{across tasks}. This couples the $T$ routers and directly targets system-level imbalance where a few experts become overloaded jointly by multiple tasks.

The overall training objective is
\begin{equation}
\mathcal{L}=\mathcal{L}_{\mathrm{task}}+\beta\,\mathcal{L}_{\mathrm{lb}},
\end{equation}
where $\beta\ge 0$ controls the regularization strength.

\subsection{Deployment and System Optimizations}
\label{sec:online}

The sparse multi-task design introduces two practical challenges for online deployment.
First, deduplicated routing leads to varying input shapes across batches, resulting in irregular execution patterns that limit the applicability of static compilation and kernel-level optimizations.
Second, such dynamic execution necessitates per-batch memory allocation.
These factors introduce additional runtime overhead and hinder the effective utilization of computational resources. We address these challenges via a custom deduplicated expert kernel and profiling-guided workspace allocator.

\subsubsection{Reindexed Grouped GEMM}
\label{sec:dedup}

The construction of the deduplicated expert set $\mathcal{U}$ inherently produces \emph{ragged tensors}, where each expert processes a varying, non-contiguous subset of the mini-batch. 
Standard dense matrix multiplications cannot efficiently support such irregular patterns. To resolve this, we propose a \textbf{Reindexed Grouped GEMM} kernel. Inspired by MegaBlocks~\cite{megablocks}, this design maps dynamic sparse activations directly into compact dense computations, bypassing variable-shape overheads. The pipeline proceeds in four phases:

\paragraph{Traffic Calculation.}
For a mini-batch of size $B$, let $b \in \{1,\dots,B\}$ index the instances and $e \in \{1,\dots,E\}$ index the experts. $\mathcal{U}^{(b)}$ denotes the set of unique experts required by instance $b$ across all tasks. We define the effective load for expert $e$ as $n_e = \sum_{b=1}^{B}\mathbb{I}[e\in\mathcal{U}^{(b)}]$. The total number of active expert executions is $N_{\mathrm{act}} = \sum_{e=1}^{E} n_e$.

\paragraph{Execution-Time Reindexing (Gather \& Map).}
Standard frameworks struggle with dynamic tensor shapes. We explicitly manage memory by gathering inputs into a static workspace.
\begin{itemize}[leftmargin=*, nosep, topsep=2pt]
    \item \textbf{Gather:} We gather hidden states $\mathbf{x}$ of instances activating expert $e$ into a compact matrix. All such matrices are packed contiguously into a global tensor $\mathbf{X} \in \mathbb{R}^{N_{\mathrm{act}} \times d_{\mathrm{in}}}$.
    \item \textbf{Index Mapping ($\pi$):} Simultaneously, we record a back-mapping $\pi$. For every activated pair $(b, e)$, $\pi(b, e)$ stores the row index in $\mathbf{X}$ corresponding to this computation.
\end{itemize}

\paragraph{Grouped GEMM}
We perform expert computations in a single batched kernel:
\begin{equation}
\mathbf{O} = \text{GroupedGEMM}(\mathbf{X}, \{\mathbf{W}_e\}, \{n_e\}),
\end{equation}
where each expert weight $\mathbf{W}_e$ is applied to its corresponding contiguous segment of $\mathbf{X}$ with size $n_e$, and $\mathbf{O} \in \mathbb{R}^{N_{\mathrm{act}} \times d_{\mathrm{out}}}$ stores the resulting expert outputs.

\paragraph{Task-wise Representation Reconstruction}
Finally, we recover task-specific representations $\mathbf{h}_t^{(b)}$ by querying the packed results using the map $\pi$:
\begin{equation}
\mathbf{h}_t^{(b)} = \sum_{e \in \mathcal{K}_t^{(b)}} p_{t,e}^{(b)} \cdot \mathbf{O}[\pi(b, e)].
\end{equation}
This ensures that if multiple tasks select the same expert for an instance, the expert is evaluated exactly once, and the result is reused via the index map.

\subsubsection{Profiling-Guided Workspace Allocation}
\label{sec:workspace}

% TODO
Due to deduplicated sparse routing, the number of expert executions $N_{\mathrm{act}}$ varies significantly across batches, leading to fluctuating workspace requirements.
Consequently, the temporary workspace required $M_{\mathrm{req}}$ for the packed inputs $\mathbf{X}$ and expert outputs $\mathbf{O}$ varies linearly:
\begin{equation}
M_{\mathrm{req}} \propto N_{\mathrm{act}} \cdot (d_{\mathrm{in}} + d_{\mathrm{out}}).
\end{equation}
% A naive strategy allocates for the worst case (i.e., $N_{\max} \approx B E$), which leads to severe memory over-provisioning and fragmentation. 

% To balance memory efficiency and latency, we implement a profiling-guided workspace allocation strategy. Based on offline profiling of the load distribution $P(N_{\mathrm{act}})$, we pre-allocate a set of workspace pools organized by discrete size classes $\{C_1 < C_2 < \dots < C_k\}$. At runtime, immediately after computing $N_{\mathrm{act}}$, we retrieve a buffer from the smallest sufficient size class $j$ such that $C_j \ge N_{\mathrm{act}}$. This approach achieves near-optimal memory footprint with negligible allocation overhead.
To balance memory efficiency and latency, we implement a profiling-guided workspace allocation strategy. Based on offline profiling of the load distribution $P(N_{\mathrm{act}})$, we pre-allocate a shared GPU memory pool composed of a set of fixed-size memory pages, where each page roughly corresponds to the memory footprint required for processing a single item. 
At runtime, after determining the required number of active pages, each execution thread allocates a contiguous block of pages from the shared pool and releases them back to the pool upon completion of expert computation.
By leveraging offline profiling to estimate the expected concurrency level, we provision the pool capacity to reduce contention-induced waiting while maximizing overall memory utilization.

% \subsubsection{Systems Optimizations for Online Serving}
% Beyond sparse routing, cross-task deduplication, reindexed grouped GEMM, and adaptive workspace allocation, SMES-Network further incorporates several critical systems optimizations to meet production latency and throughput requirements:
% \begin{itemize}[leftmargin=*, nosep]
%      \item \textbf{Operator fusion.} We fuse common sequences of linear, normalization, and activation operations within expert and gating networks into single kernel, alleviating memory bandwidth pressure and reducing kernel launch overhead.
%      \item \textbf{Mixed precision inference.} We employ a BF16-training and FP16-inference strategy, which reduces memory bandwidth and maximizes GPU throughput without sacrificing accuracy.
%      \item \textbf{Kernel-aware layout design.} The data layout of instance and expert activations is chosen to align with the underlying GEMM~\cite{cutlass} implementation for grouped sparse matmul, further improving efficiency.
% \end{itemize}

% These optimizations are complementary to the algorithmic design of SMES-Network and are essential for deploying massive-scale sparse MoE recommendation models under strict online latency constraints.\dk{Is there data on the effect/impact of each optimization} \zyk{delete?}

% TODO
\subsection{Discussion}
\label{sec:complexity}

We discuss the computational complexity and scalability of SMES against the standard dense MMoE variant.
\begin{itemize}[leftmargin=*, nosep, topsep=2pt, itemsep=3pt]
    \item \textbf{Dense MMoE:} In standard MMoE, every instance is processed by all $E$ experts. Consequently, both FLOPs and memory costs scale linearly with the number of experts $\mathcal{O}(E)$, making large-scale deployment increasingly expensive.
    \item \textbf{SMES:}
    SMES decouples inference cost from the total model capacity.
    With sparse routing and deduplicated execution (Sec.~\ref{sec:union}), inference only evaluates the set of unique activated experts $\mathcal{U}$, resulting in a computational complexity of $\mathcal{O}(|\mathcal{U}|)$.
    Since $|\mathcal{U}| \ll E$ due to the task-shared expert set $\mathcal{S}$, the inference cost is significantly lower than $\mathcal{O}(E)$, ensuring both sparsity and scalability as the number of experts expands.
    Meanwhile, the proposed multi-task load-balancing regularization (Sec.~\ref{sec:balance}) promotes balanced expert utilization under sparse activation, maintaining training stability.
    Finally, our online optimization strategies (Sec.~\ref{sec:online}) further ensure that the sparse design translates into bounded and stable serving latency in real-world deployments.
\end{itemize}\section{Experiments}
\label{sec:Experiment}

\subsection{Experiment Settings}
\label{sec: Experiment Settings}

\subsubsection{Datasets}
\label{sec:Datasets}

We adopt KuaiRand-1K a benchmark recommendation dataset (detailed in Table~\ref{tab:dataset_details}), as the public dataset for offline experiments, and simultaneously conduct large-scale industrial offline studies on the Kuaishou short-video recommendation platform. To maintain consistent terminology, the valid\_play label in KuaiRand-1K is mapped to Effective-view. The industrial dataset comprises 400 million users and generates 50 billion daily interaction logs. Each sample corresponds to a user-video pair and the associated feedback, representing a recommendation interaction.
To align with practical industrial requirements, we focus on four key prediction tasks: \textbf{Effective-view}, \textbf{Long-view}, \textbf{Click}, and \textbf{Like}.

\begin{table}[t]
  \centering
  \caption{Dataset of public KuaiRand and industrial Kuaishou}
  \label{tab:dataset_details}
  \resizebox{\columnwidth}{!}{%
  \begin{tabular}{lccccc}
    \toprule
    Dataset & Users & Items & Instances & Features & \#Tasks \\
    \midrule
    KuaiRand-1K & 1,000 & 4,369,953 & $1.17 \times 10^7$ & 92 & 12 \\
    Kuaishou & $4 \times 10^9$ & $3 \times 10^8$ & $5 \times 10^{10}$ & 326 & 21 \\
    \bottomrule
  \end{tabular}}
\end{table}

\begin{table*}
  \centering
  \caption{Offline results on public KuaiRand-1K and Kuaishou industrial datasets. Boldface denotes the highest score and underline indicates the best result of the baselines.}
  \label{tab:Offline_results_and_Online_A/B_testing}
  \resizebox{\textwidth}{!}{%
  \begin{tabular}{lccccccccc|ccccccccc}
    \toprule
    \multicolumn{10}{c}{KuaiRand-1K Dataset} & \multicolumn{9}{c}{Kuaishou Industrial Dataset} \\
    \cmidrule(lr){2-10} \cmidrule(lr){11-19}
    \multirow{2}{*}{Model} & \multicolumn{2}{c}{Effective-view} & 
    \multicolumn{2}{c}{Like} & \multicolumn{2}{c}{Follow} & \multicolumn{2}{c}{Comment} & \multirow{2}{*}{\#Params} & \multicolumn{2}{c}{Effective-view} & 
    \multicolumn{2}{c}{Long-view} & \multicolumn{2}{c}{Click} & \multicolumn{2}{c}{Like} & \multirow{2}{*}{\#Params} \\
    \cmidrule(lr){2-3} \cmidrule(lr){4-5} \cmidrule(lr){6-7} \cmidrule(lr){8-9} \cmidrule(lr){11-12} \cmidrule(lr){13-14} \cmidrule(lr){15-16} \cmidrule(lr){17-18}
    & AUC & GAUC & AUC & GAUC & AUC & GAUC & AUC & GAUC & & AUC & GAUC & AUC & GAUC & AUC & GAUC & AUC & GAUC & \\
    \midrule
    MMoE & 0.7657 & 0.6355 & 0.8586 & 0.6119 & 0.8004 & 0.6188 & 0.7855 & 0.5957 & 386.84M & 0.7867 & 0.7319 & 0.8223 & 0.7645 & \underline{0.7370} & 0.6663 & 0.9650 & 0.8479 & 243.96M \\
    PLE & 0.7670 & 0.6363 & \underline{0.8939} & \underline{0.6280} & 0.7989 & \underline{0.6220} & 0.8274 & 0.5990 & 387.79M & 0.7875 & 0.7325 & 0.8220 & 0.7640 & 0.7363 & 0.6659 & 0.9647 & 0.8476 & 251.16M \\
    HoME & 0.7672 & 0.6417 & 0.8891 & 0.6251 & 0.7973 & 0.5997 & 0.7994 & \underline{0.6031} & 387.31M & 0.7876 & 0.7332 & 0.8204 & 0.7626 & 0.7322 & 0.6664 & 0.9650 & 0.8477 & 269.66M \\
    MoME & 0.7667 & 0.6313 & 0.8384 & 0.6201 & 0.8033 & 0.5922 & 0.7988 & 0.5943 & 389.68M & 0.7866 & 0.7319 & 0.8220 & 0.7641 & 0.7367 & 0.6660 & 0.9649 & 0.8478 & 271.47M \\
    Rankmixer & \underline{0.7699} & \underline{0.6419} & 0.8818 & 0.6230 & \underline{0.8060} & 0.5975 & \underline{0.8301} & 0.5932 & 387.31M & \underline{0.7890} & \underline{0.7342} & \underline{0.8234} & \underline{0.7658} & 0.7295 & \underline{0.6665} & \underline{0.9651} & \underline{0.8488} & 280.90M \\
    \midrule
    SMES-S & \textbf{0.7722} & \textbf{0.6426} & \textbf{0.9102} & \textbf{0.6284} & \textbf{0.8075} & \textbf{0.6224} & \textbf{0.8465} & \textbf{0.6046} & 386.13M & \textbf{0.7895} & \textbf{0.7356} & \textbf{0.8235} & \textbf{0.7663} & \textbf{0.7419} & \textbf{0.6690} & \textbf{0.9655} & \textbf{0.8492} & 272.70M \\
    Improved & +0.23\% & +0.07\% & +1.63\% & +0.04\% & +0.15\% & +0.04\% & +1.64\% & +0.15\% & - & +0.05\% & +0.14\% & +0.01\% & +0.05\% & +0.49\% & +0.25\% & +0.04\% & +0.04\% & - \\
    \midrule
    SMES-L & \textbf{0.7741} & \textbf{0.6448} & \textbf{0.9182} & \textbf{0.6309} & \textbf{0.8159} & \textbf{0.6251} & \textbf{0.8502} & \textbf{0.6052} & 510M & \textbf{0.7918} & \textbf{0.7371} & \textbf{0.8241} & \textbf{0.7671} & \textbf{0.7424} & \textbf{0.6712} & \textbf{0.9663} & \textbf{0.8499} & 633M \\
    Improved & +0.42\% & +0.22\% & +2.43\% & +0.29\% & +0.99\% & +0.31\% & +2.01\% & +0.21\% & - & +0.28\% & +0.29\% & +0.07\% & +0.13\% & +0.54\% & +0.47\% & +0.12\% & +0.11\% & - \\
    \bottomrule
  \end{tabular}}
% \vspace{2pt}
\caption*{
\footnotesize
\raggedright
SMES-S and SMES-L denote two scaled configurations of SMES.
SMES-S matches the parameter budget of the baselines,
whereas SMES-L increases expert capacity under a fixed sparsity ratio of 8\% to explore the performance ceiling.
}
\vspace{-6pt}
\end{table*}

\subsubsection{Baselines}
\label{sec:Baselines}
We compare SMES with representative multi-task learning methods: 
\textbf{MMOE}, the standard multi-expert framework for recommendation; 
\textbf{PLE}, which alleviates the seesaw phenomenon by separating shared and task-specific components; 
\textbf{MoME}, which applies parameter sparsification to balance performance and resource usage; 
and \textbf{HoME}, a MoE-based framework successfully deployed in large-scale industrial recommendation systems using hierarchical masking for stability. 
We also evaluate \textbf{RankMixer}, which leverages sparse MoE for parameter scaling.

\subsubsection{Evaluation Metrics}
\label{sec: Evaluation Metrics}
We evaluate our model using two widely adopted ranking metrics: \textbf{AUC} and \textbf{GAUC}, with GAUC being the primary offline metric in our short-video recommendation service \cite{zhou2018deep}. GAUC computes the AUC independently for each user and aggregates them in a weighted manner:
\[
\text{GAUC} = \sum_{u} w_u \cdot \text{AUC}_u, \quad \text{where} \quad w_u = \frac{\#\text{samples}_u}{\sum_i \#\text{samples}_i}
\]
where $w_u$ denotes the user's sample proportion.

\subsection{Offline Experiments}
\label{sec:Offline Experiments}

This section validates the effectiveness and generalization of SMES on recommendation tasks. Table ~\ref{tab:Offline_results_and_Online_A/B_testing} presents the offline evaluation results on two real-world datasets: KuaiRand-1K and Kuaishou, with models grouped by similar parameter budgets.

As shown in Table~\ref{tab:Offline_results_and_Online_A/B_testing}, PLE consistently outperforms MMoE on most tasks, indicating that explicitly separating shared and task-specific components is beneficial for mitigating inter-task interference.
In contrast, MoME shows relatively weaker performance across tasks, suggesting that naive parameter sparsification may compromise representation capacity.
RankMixer, which emphasizes scalable sparse modeling, achieves competitive results and outperforms most baselines. 
SMES-S, a configuration of SMES designed to match the parameter budgets of representative baselines, delivers superior performance across all tasks on both public and industrial datasets. In particular, when compared to RankMixer, SMES-S obtains an average GAUC gain of around 0.1\% on watching-time tasks (e.g., Effective-view), while delivering substantial gains on interaction tasks (e.g., Click, Like). These results highlight that SMES effectively addresses the misalignment between uniform capacity scaling and task-specific supervision.
To further investigate the scalability of SMES, we scale up the model’s parameters while maintaining a comparable number of activated parameters to the baselines (i.e., variant SMES-L with expert sparsity of 8\%).
As shown in the results, SMES-L brings further improvements over SMES-S on all tasks, demonstrating that SMES can effectively leverage additional model capacity to improve performance. Together with the analysis in Section~\ref{sec:Scalability and Lantency}, these findings indicate that SMES yields significant gains without incurring additional latency overhead.

\subsection{Scalability Analysis}
\label{sec:Scalability and Lantency}
\begin{figure}[t] 
	\centering  
	\includegraphics[width=0.48\textwidth]{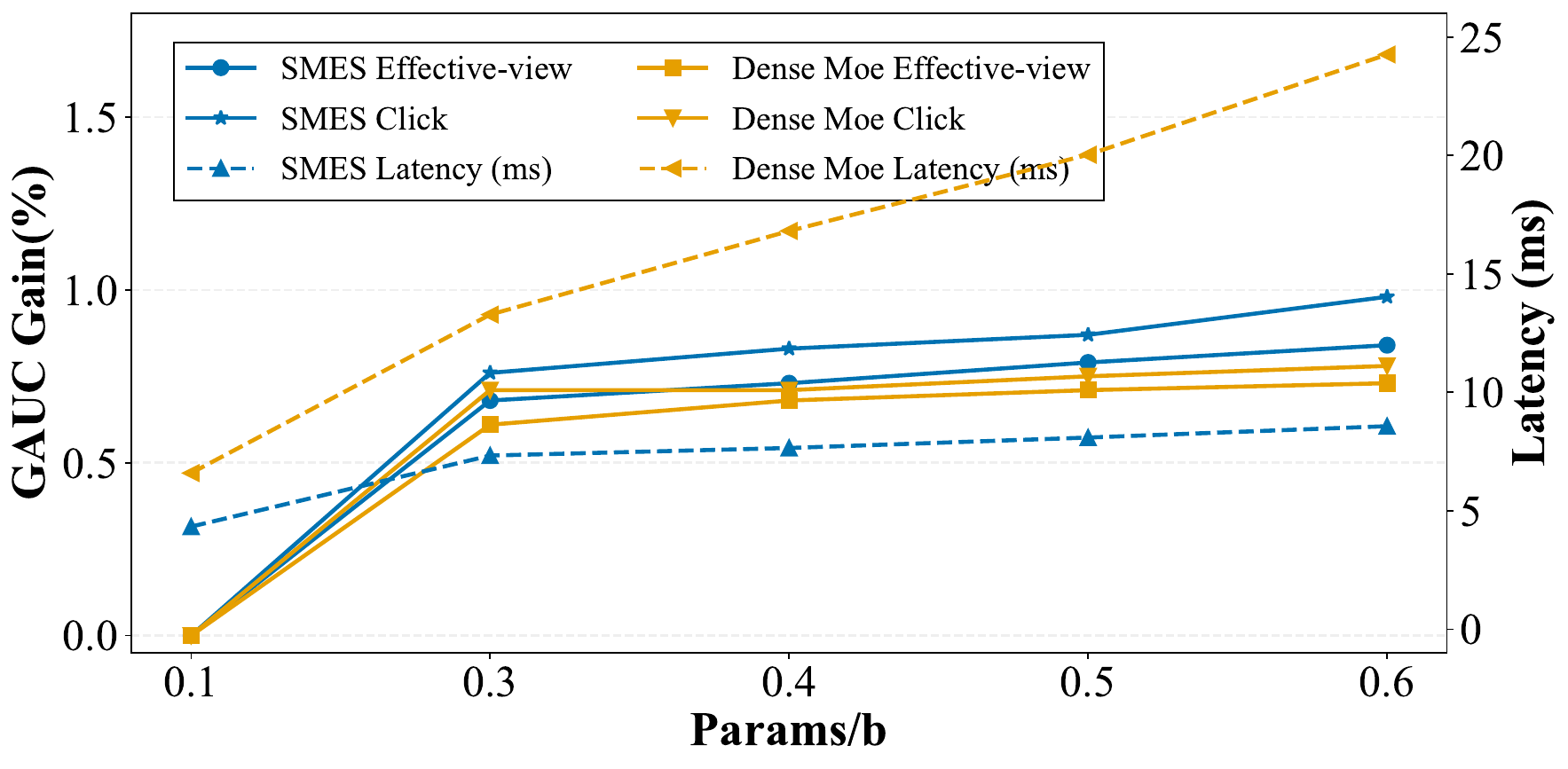}
	\caption{\textbf{Scalability Analysis of SMES vs. Dense MoE: Performance and Latency in Effective View and Click Tasks.} }
	\label{scale}
	\vspace{-6pt}
\end{figure}

This section presents the scaling curves for SMES and dense models with respect to parameter size, systematically evaluating scalability, task-adaptive capacity allocation, and latency efficiency. As shown in Figure ~\ref{scale}, SMES consistently benefits from parameter scaling across heterogeneous tasks, effectively overcoming the diminishing returns observed in dense models. Its sparse structure and dynamic expert activation mechanism enable task-adaptive capacity allocation, providing superior performance relative to the dense baseline.
In contrast, SMES maintains stable inference latency even as model parameters increase, satisfying the strict online serving constraints of industrial recommender systems. For instance, expanding SMES to 0.6B parameters results in an average GAUC improvement of 0.84\% while incurring only a 4ms latency increase. However, dense models exhibit nearly linear latency growth with scale, limiting their practical deployment. These results validate that SMES balances parameter scaling, task specialization, and latency, establishing a practical paradigm for multi-task recommendation.

\subsection{Ablation Study}
\label{sec:Ablation Experiment}

\begin{table}[t]
  \centering
  \caption{Ablation Study of SMES Components with GAUC.}
  \label{tab:ablation_combined_transposed}
  \resizebox{\columnwidth}{!}{%
  \begin{tabular}{l|l|cccccc}
    \toprule
    Dataset & Task & \textbf{ours} & \textbf{\textit{w/o}} Task-SHA & \textbf{\textit{w/o}} Task-ADA & \textbf{\textit{w/o}} Reg \\
    \midrule
    \multirow{4}{*}{KuaiRand-1K} 
    & Effective-view & \textbf{0.6426} & 0.6342 & 0.6409 & 0.6422 \\
    & Like           & \textbf{0.6284} & 0.6017 & 0.6010 & 0.6135 \\
    & Follow         & \textbf{0.6224} & 0.6198 & 0.6144 & 0.6174 \\
    & Comment        & \textbf{0.6046} & 0.5891 & 0.6045 & 0.5956 \\
    \midrule
    \multirow{4}{*}{Kuaishou} 
    & Effective-view & \textbf{0.7356} & 0.7344 & 0.7342 & 0.7345 \\
    & Long-view      & \textbf{0.7663} & 0.7650 & 0.7648 & 0.7649 \\
    & Click          & \textbf{0.6690} & 0.6670 & 0.6666 & 0.6681 \\
    & Like           & \textbf{0.8492} & 0.8484 & 0.8481 & 0.8483 \\
    \bottomrule
  \end{tabular}}
\end{table}

 We conduct ablation experiments on the public KuaiRand-1K and the Kuaishou industrial datasets to validate the effectiveness of three core SMES components, namely Task-shared experts (\textbf{\textit{w/o}} Task-SHA), Task-adaptive experts (\textbf{\textit{w/o}} Task-ADA), and load balancing regularization  (\textbf{\textit{w/o}} Reg). The full SMES model attains better GAUC performance than all ablation variants across all tasks on both datasets, which verifies the synergistic effect of its core components. Removing Task-SHA leads to significant performance degradation. This aligns with our method, which prioritizes the selection of shared experts from the expert pool to capture cross-task patterns. Removing Task-ADA also causes notable performance drops, highlighting the necessity of task-adaptive capacity allocation for intra-task specialization. Removing load-balancing regularization leads to a moderate performance drop. This indicates that the regularizer plays an important role in mitigating expert overload by coordinating the aggregated activation across all tasks.

\subsection{Hyper-Parameter Sensitivity}
\label{sec:Hyper-Parameter}

\begin{figure}[t] 
    \centering  
    \subcaptionbox{Sensitivity analysis of hidden dimensions.\label{fig:hyper_dim_sub}}{%
        \includegraphics[height=2.7cm, width=0.4\textwidth]{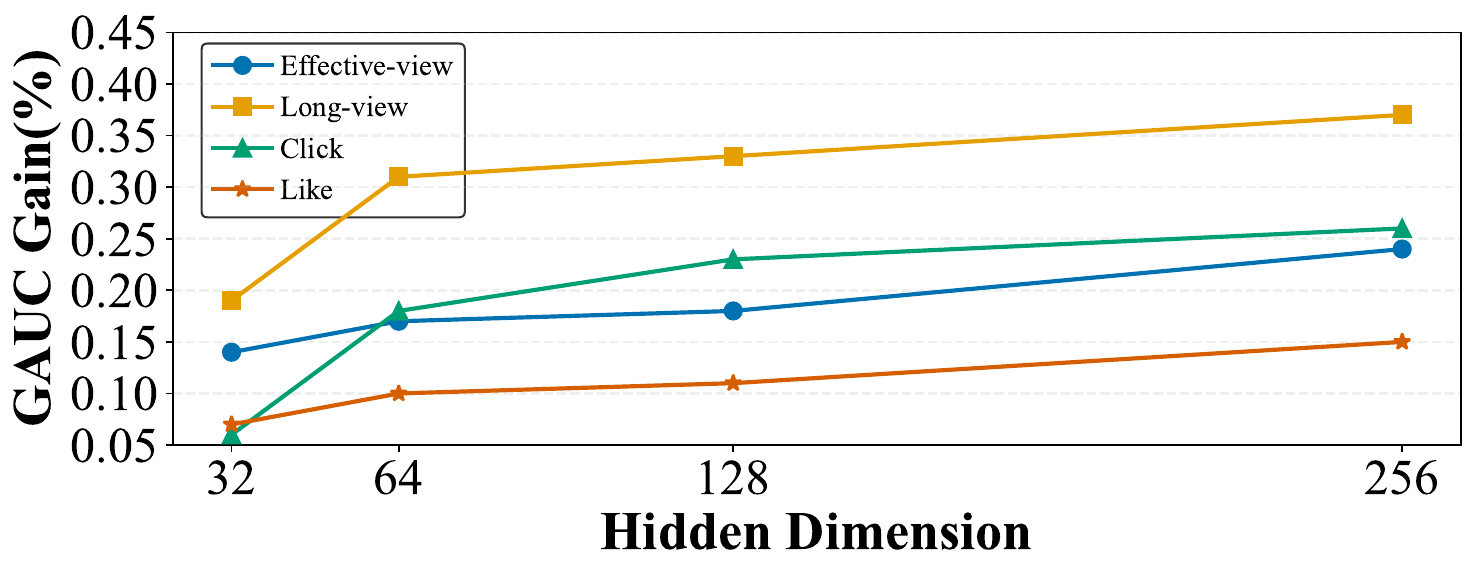}%
    }
    \hspace{5pt}
    \subcaptionbox{Sensitivity analysis of expert numbers.\label{fig:hyper_expert_sub}}{%
        \includegraphics[height=2.7cm, width=0.4\textwidth]{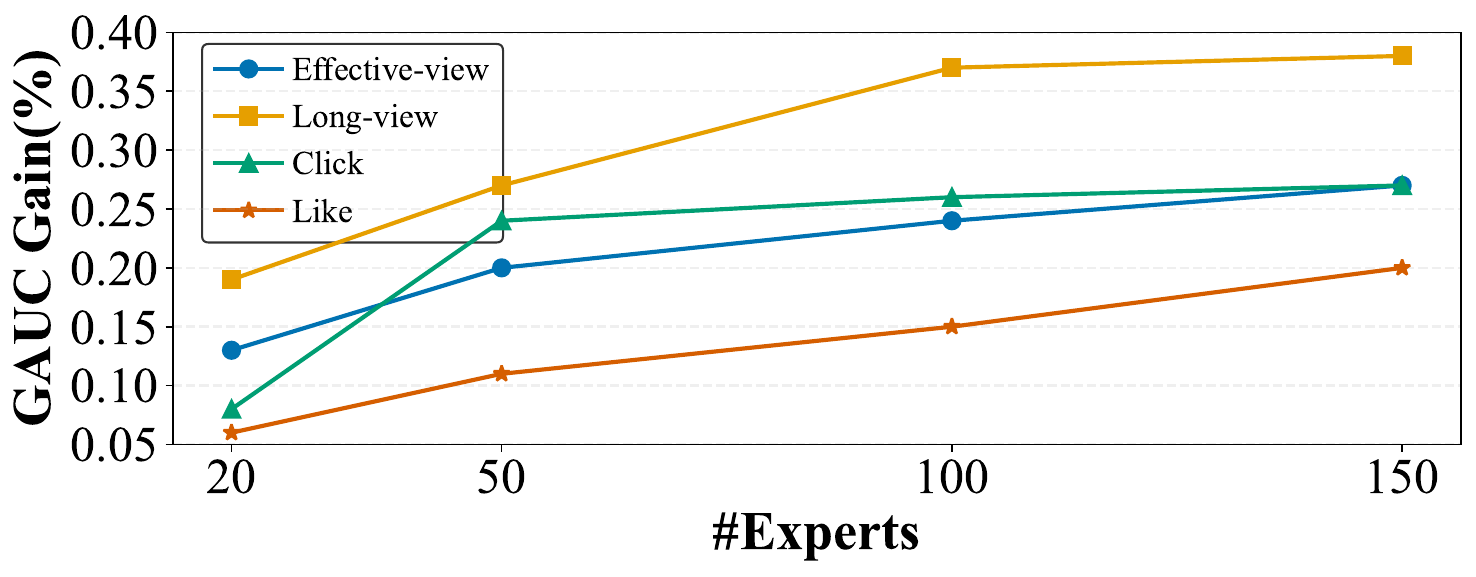}%
    }
    \caption{Hyper-Parameter Sensitivity of SMES. We evaluate the impact of hidden dimensions and expert numbers on SMES performance across core recommendation tasks (Effective-view, Like, Follow, Comment).}
    \label{fig:hyper_param}
    \vspace{-6pt} 
\end{figure}
We examine the hyperparameter sensitivity of SMES with respect to the hidden dimension size and the number of experts to evaluate the model’s robustness and scalability. For the hidden dimension analysis, we fix the number of experts at 100 and evaluate four distinct dimension settings. For the expert count analysis, we fix the hidden dimension at 256 and assess four different settings for the number of experts.
Experimental results in Figure~\ref{fig:hyper_dim_sub} reveal a positive correlation between hidden dimension and model performance: the prediction accuracy exhibits a steady upward trend as the hidden dimension expands, while core metrics also demonstrate continuous and statistically meaningful improvements. From Figure~\ref{fig:hyper_expert_sub}, it can be observed that model performance exhibits an overall upward trend with increasing expert count, where the GAUC continuously improves and reaches its optimal value at the maximum expert count. 
Hidden dimension expansion delivers relatively modest improvements overall, whereas increasing the number of experts brings more pronounced performance gains by comparison. 
Furthermore, different tasks exhibit varying sensitivities to parameter adjustments, with the Long-view task showing the strongest response to such changes across all tasks.

% This section investigates the hyperparameter sensitivity of SMES with respect to the hidden representation dimension and the expert pool size, aiming to systematically evaluate the framework’s robustness, task-adaptive capacity allocation, and performance scalability. For the hidden dimension analysis, we fix the expert pool size to 100 and conduct experiments across four distinct hidden dimension settings. For the expert pool analysis, we fix the hidden dimension to 256 and evaluate four different expert pool sizes.

\subsection{Visualization of Task–Expert Activation}
\label{sec:Situation}

\begin{figure}[t] 
    \centering  
    \includegraphics[width=0.49\textwidth]{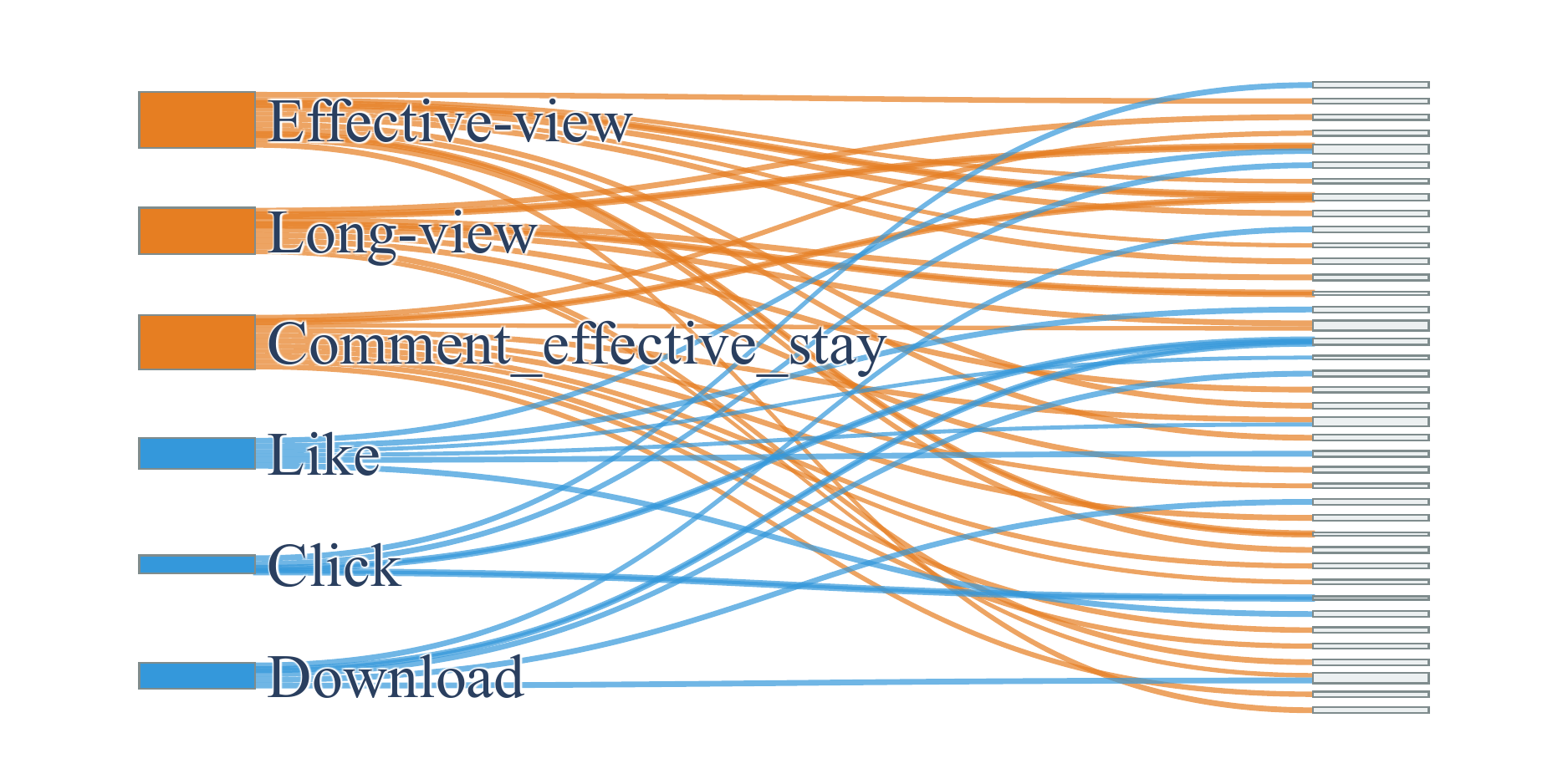}
    \caption{\textbf{Task-Expert Activation of our SMES.} For the sake of clarity and conciseness, we only display three most-activated with a high number of assigned experts, marked in orange, and three least-activated tasks with a low number, marked in blue.}
    \label{expert-task}
    \vspace{-6pt}
\end{figure}

To verify the effectiveness of adaptive expert selection across distinct tasks, we select the three most-activated and three least-activated tasks and extract the experts they activate, as shown in Figure~\ref{expert-task}. It can be observed that the number of activated experts for each task is adaptively matched to its data scale: dense tasks activate more experts for superior preference modeling, while sparse tasks use fewer experts to prevent over-parameterization. 
Meanwhile, the balanced activation distribution across all tasks vividly demonstrates the exceptional effectiveness of the load-balancing regularization in SMES, which guarantees efficient and balanced expert utilization to underpin the scalable performance.
% of SMES in multi-task recommendation.

\subsection{Online A/B Test}
\label{sec:Online A/B Test}

To evaluate the industrial feasibility of SMES, we conducted online A/B tests on the Kuaishou Single Page with 3.5\% of production traffic from October 25 to October 31, 2025. To ensure the fairness of the A/B test, we only replaced HoME ~\cite{home} in the baseline with the SMES module while keeping other components and hyperparameters unchanged. 
Online A/B test results validate that SMES achieves statistically significant improvements across all business-critical metrics, outperforming the baseline by a substantial margin. In the Kuaishou Single Page scenario, SMES delivers an average user watch time lift of +0.31\%. Core interaction metrics also obtain notable and statistically meaningful gains, including Like +0.64\%, Follow +1.56\%, Comment +2.45\%. 
SMES mitigates the trade-off between parameter scaling and inference latency, reducing inference latency by 50\% relative to the dense MoE baseline with comparable capacity. SMES has been fully deployed in Kuaishou’s large-scale production environment, serving over 400 million users daily.

\section{Conclusion}
\label{sec: Conclusion}
We present SMES, a scalable sparse MoE framework for multi-task recommendation that addresses the misalignment between uniform parameter scaling and heterogeneous task capacities. SMES uses progressive expert routing, splitting activation into a task-shared subset for universal patterns and task-adaptive experts for specialized modeling, thereby bounding the number of active experts per instance. A global load-balancing regularizer mitigates cross-task expert overload, and optimized matrix computation with efficient memory allocation enables low-latency deployment. 
Extensive experiments on public benchmarks and large-scale industrial datasets demonstrate that SMES consistently surpasses dense and naive sparse MoE baselines, achieving substantial online gains under strict latency budgets.

% \begin{acks}

% \end{acks}

\bibliographystyle{ACM-Reference-Format}
\bibliography{main-SMES}

\appendix

\end{document}